\documentclass[%
 reprint,
showpacs,preprintnumbers,
nofootinbib,
 amsmath,amssymb,
 aps,
 showkeys,
 showpacs
]{revtex4-1}
\usepackage{graphics}
\usepackage{bbold}
\usepackage{color}
\usepackage{amsmath}
\usepackage{graphicx}
\usepackage{subfigure}

\usepackage{lipsum}

\definecolor{simone}{RGB}{255,0,0}
\definecolor{darkblue}{RGB}{0,0,150}
\def\gp  		{\mbox{$G$-parity}}
\def\jp  		{\ensuremath{{J/\psi}}}
\def\pipi  		{\ensuremath{\pi^+\pi^-}}
\def\ee  		{\ensuremath{e^+ e^-}}
\def\ggga 		{\ensuremath{gg\gamma}}
\def\a	 		{\ensuremath{\mathcal{A}}}
\def\ag 		{\ensuremath{\mathcal{A}_\gamma}}
\def\agg 		{\ensuremath{\mathcal{A}_{gg\gamma}}}
\def\aggg 		{\ensuremath{\mathcal{A}_{ggg}}}
\def\b 			{\ensuremath{\mathcal{B}}}
\def\bg 		{\ensuremath{\mathcal{B}_\gamma}}
\def\bgg 		{\ensuremath{\mathcal{B}_{gg\gamma}}}
\def\bggg 		{\ensuremath{\mathcal{B}_{ggg}}}
\def\B 			{\ensuremath{\mathcal{B}}}
\def\be			{\begin{eqnarray}}
\def\en			{\end{eqnarray}}
\def\nen		{\nonumber\end{eqnarray}}
\def\no			{\nonumber}
\def\hh			{\hspace{2mm}}

\def\re			{\ensuremath{{\rm Re}}}
\def\im			{\ensuremath{{\rm Im}}}
\def\epg 		{\ensuremath{{\eta'\gamma}}}
\def\ds			{\displaystyle}

\def\R        {\ensuremath{\mathbb{R}}}

\def\nuovo	  {\color{black}}
\def\lq			{\ensuremath{\left[}}
\def\rq			{\ensuremath{\right]}}
\def\lt			{\ensuremath{\left(}}
\def\rt			{\ensuremath{\right)}}
\def\f		{\ensuremath{f_1(1285)}}
\def\eg 		{\ensuremath{{\eta\gamma}}}
\def\fg 		{\ensuremath{{f_1\gamma}}}
\newcommand{\hl}[1]{{#1}}
\DeclareMathOperator*{\SumInt}{%
\mathchoice%
  {\ooalign{$\displaystyle\sum$\cr\hidewidth$\displaystyle\int$\hidewidth\cr}}
  {\ooalign{\raisebox{.14\height}{\scalebox{.7}{$\textstyle\sum$}}\cr\hidewidth$\textstyle\int$\hidewidth\cr}}
  {\ooalign{\raisebox{.2\height}{\scalebox{.6}{$\scriptstyle\sum$}}\cr$\scriptstyle\int$\cr}}
  {\ooalign{\raisebox{.2\height}{\scalebox{.6}{$\scriptstyle\sum$}}\cr$\scriptstyle\int$\cr}}
}
\begin{document}
\title{$G$-parity violating amplitudes in the $J/\psi \to \pi^+ \pi^-$ decay}
\author{Rinaldo Baldini Ferroli}
\affiliation{%
Laboratori Nazionali di Frascati dell'INFN, Frascati, Italy}%
\author{Alessio Mangoni}
\affiliation{%
Dipartimento di Fisica e Geologia, Universit\`a degli Studi di Perugia and INFN Sezione di Perugia, Perugia, Italy}%
\author{Simone Pacetti}
 \email{simone.pacetti@pg.infn.it}
\affiliation{%
Dipartimento di Fisica e Geologia, Universit\`a degli Studi di Perugia and INFN Sezione di Perugia, Perugia, Italy}%
%
%
\begin{abstract}%
The decays of the negative \gp\ meson \jp\ into even numbers of pions violate \gp. Such decays, as well as other \jp\ decays into hadrons, can be parametrized in terms of three main intermediate virtual states: \hl{one photon, one photon plus two gluons, and three gluons}. Since the electromagnetic interaction does not conserve $G$-parity, \jp\ decays into positive \gp\ final states should be dominantly electromagnetic. Nevertheless, the one-photon contribution to $\jp\to\pipi$, that can be estimated by exploiting the cross section $\sigma(\ee\to\pipi)$, differs from the observed decay probability for \hl{more than} 4.5 standard deviations.
\\
We present a computation of the \ggga\ amplitude based on a phenomenological
description of the decay mechanism in terms of dominant intermediate states \hl{$\eg$, $\epg$ and $\f\gamma$}. The obtained value is of the order of the electromagnetic contribution.
\end{abstract} 
\maketitle
\section{Introduction}
\label{intro}
\begin{figure}[h!]
\begin{flushright}
            \includegraphics[width=0.45\textwidth]{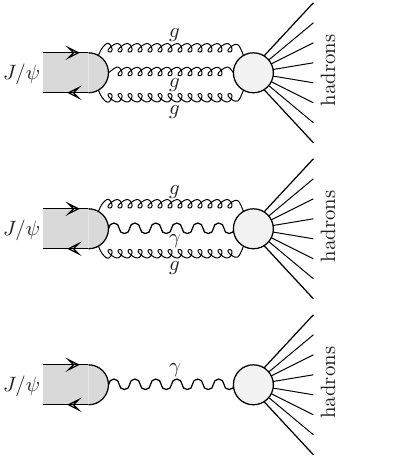}%
\caption{The three intermediate states: $ggg$, $gg \gamma$ and $\gamma$.}
\label{fig:3.states.h}
\end{flushright}
\end{figure}

The $J/\psi$ meson, having negative $C$-parity,  $C_{\jp} = -1$, and isospin $I_{\jp}=0$, has negative \gp\, $G_{\jp}$, being $G_{\jp} = C_{\jp} \cdot (-1)^{I_{\jp}}=-1$. \gp\ is a multiplicative quantum number so that a set of $n$ pions has $G_{n \pi} = (-1)^n$. Therefore, the decays $\jp\to 2n\,\pi$, with $n=1,2,\ldots$, do not conserve \gp.
\\
Strong interaction preserves \gp\ as a consequence of its charge conjugation and isospin conservation. Electromagnetic and weak interactions can violate \gp, being not invariant under $G$ transformations.
\\
In general the decay $J/\psi \to n \,\pi$ is parametrized in terms of three main intermediate states~\cite{Kopke:1988cs}: three gluons, $ggg$ (purely strong); two gluons plus one photon, $gg \gamma$ (mixed); one photon, $\gamma$ (purely electromagnetic). The corresponding Feynman diagrams are shown in Fig.~\ref{fig:3.states.h}.
\\
The contribution of the intermediate state with three photons, that has the same structure of the three-gluon one, is neglected being of order $\alpha^2$ with respect to that with a single photon.
\\
 The total decay amplitude can be written as the sum of the three contributions: \aggg, \agg, \ag, corresponding to the three mechanisms represented in Fig.~\ref{fig:3.states.h}, and that, as a consequence of strengthen of the underlying interactions, follow the hierarchy: $|\aggg|\gg|\agg|\simeq|\ag|$.\\
In light of that, the branching ratio (BR) of \jp\ decay into a hadronic final state, $h$, is decomposed as
\be
\mathcal B(h) = \bggg(h) + \bgg(h) + \bg(h) + \mathcal{I}(h) \,,
\label{eq:br-contributions}
\en
where $\mathcal{B}_{X}(h)\propto |\mathcal{A}_{X}(h)|^2$, with $X=ggg$, $gg\gamma$, $\gamma$, while $\mathcal{I}(h)$ accounts for the interference terms.  
%
%
\\
\begin{figure}
\centering
\resizebox{0.45\textwidth}{!}{%
  \includegraphics{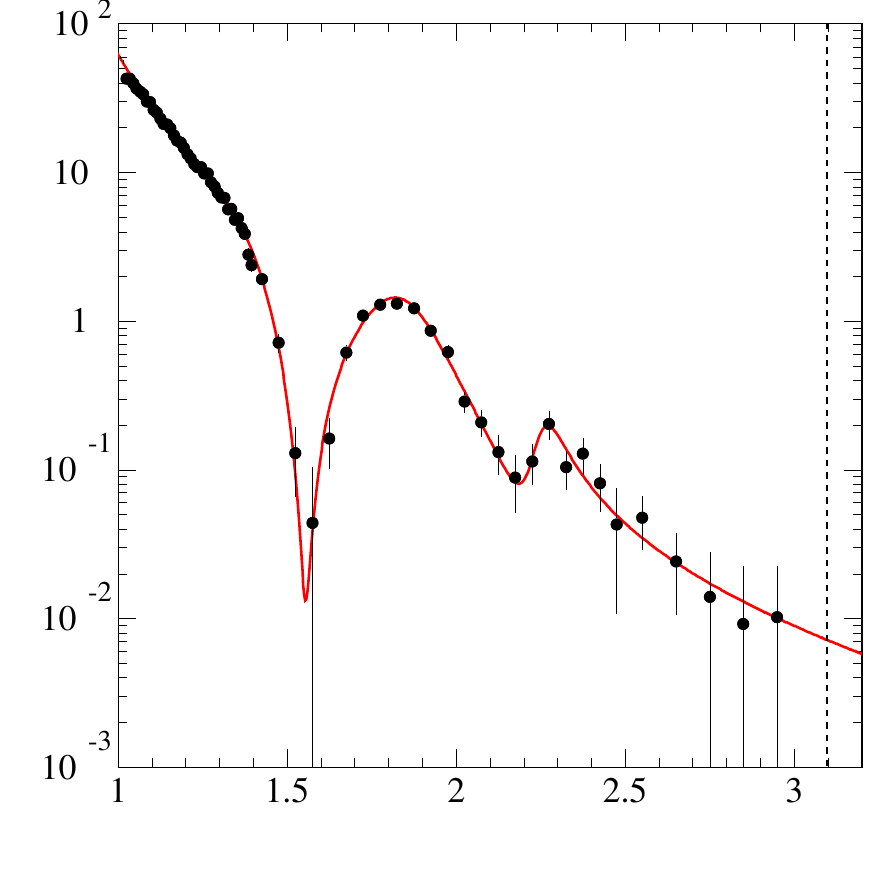}}
\put(-53,7){$\sqrt{q^2} \ (\rm GeV)$}
\put(-233,134){\rotatebox{90}{$\sigma(e^+e^- \to \pi^+\pi^-) \ (\rm nb)$}}
\caption{\textsc{BaBar} data on the $e^+e^- \to \pi^+\pi^-$ cross section and the fit (red line) from~\cite{Lees:2012cj}. The vertical dashed line shows the \jp\ mass.}
\label{fig:babar}
\end{figure}\\
In case of purely pionic final states, $h=n\,\pi$, being \gp\ preserved by the strong interaction, one expects
\be
\mathcal B(n\,\pi)\simeq
\left\{\begin{array}{lcl}
\mathcal B_\gamma(n\,\pi)  &\,\,\,& n\,\,\mbox{even}\\
	&&\\
\bggg\big(n\,\pi\big) && n\,\,\mbox{odd}\\	
\end{array}
\right.
   \,.
\nen
{\nuovo%
In fact when a decay violates isospin the purely strong amplitude \aggg\ is suppressed by the small dimensionless factor $|m_u - m_d|/\sqrt{q^2}$, where $q^2$ is the typical square momentum in the process and, $m_u$ and $m_d$ are the masses of $u$ and $d$ quarks. 
In these cases the decay proceeds through the purely electromagnetic channel.
On the contrary, the three-gluon mechanism dominates in those decays that preserve \gp.}
\\
The contribution $\bg(h)$, i.e., the BR corresponding to the third Feynman diagram of Fig.~\ref{fig:3.states.h}, can be computed in terms of the {\nuovo dressed $\ee\to h$ and bare $\ee\to\mu^+\mu^-$ cross sections, evaluated at the \jp\ mass, as~\cite{Milana:1993wk,Ferroli:2016zqf}
\be
\bg(h)= \B(\mu^+ \mu^-) \, {\sigma(\ee \to h) \over \sigma^0(\ee \to \mu^+ \mu^-)} \Bigg|_{q^2 = M_{\jp}^2}\,,
\label{eq:b-gamma}
\en
where $\B(\mu^+ \mu^-)$ is the BR of the decay $\jp\to\mu^+ \mu^-$ and $\sigma^0$ stands for the bare cross section, i.e., the cross section corrected for the vacuum-polarization contributions. 
\hl{ A detailed derivation of the formula of Eq.~\eqref{eq:b-gamma} is reported in App.~\ref{app:master-formula}.}}
In the following, if not differently specified, the adjective "dressed" will be understood, so that with "cross section" we will mean "dressed cross section".\\
The validity of the hypothesis of $\bg(h)$-dominance in the \jp\ decays that violate \gp\ can be verified for all the hadronic final states $h$, for which are also available data on the total cross section $\sigma(\ee\to h)$ at the \jp\ mass.
\\
Final states with even numbers of pions represent valuable examples, being available data on the cross sections $\sigma\big(\ee\to 2(\pipi)\big)$~\cite{babar-4pi}, $\sigma\big(\ee\to 2(\pipi\pi^0)\big)$ and $\sigma\big(\ee\to 3(\pipi)\big)$~\cite{babar-6pi} around the \jp\ mass.
\\
 The corresponding BRs $\bg(h)$, extracted from those data using Eq.~\eqref{eq:b-gamma}, agree with the total BRs from the Particle Data Group (PDG)~\cite{pdg}, i.e.,
 \be
 \bg(h)\simeq \b_{\rm PDG}(h)\,,\,
 h=2(\pipi)\,, 2(\pipi\pi^0)\,, 3(\pipi)\,.
 \nen
 {\nuovo All these decays are examined and discussed in Ref.~\cite{Ferroli:2016zqf}}. For the two-pion decay $J/\psi \to \pipi$, using the value of the cross section $\sigma(\ee\to\pipi)$ at the \jp\ mass, extrapolated from the \textsc{BaBar} data~\cite{Lees:2012cj} with a fit based on the Gounaris-Sakurai formula~\cite{Gounaris:1968mw}, data and fit are shown in Fig.~\ref{fig:babar}, the BR due to the one-photon exchange mechanism is
{\nuovo\be
\bg(\pipi)=(4.7\pm 1.7)\times 10^{-5}\,,
\label{eq:bg-pipi}
\en}%
to be compared with~\cite{pdg}
\be
\b_{\rm PDG}(\pipi) = (14.7 \pm 1.4) \times 10^{-5}\,.
\label{eq:bpdg-pipi}
\en
\hl{(See for instance Refs.~\cite{Czyz:2009vj,Seth:2012nn} and references therein for other parametrizations of pion form factors the neighbourhood of the \jp\ resonance).}
In the case of $\pi^+ \pi^-$ final state the purely electromagnetic BR, Eq.~\eqref{eq:bg-pipi}, differs from the PDG value, Eq.~\eqref{eq:bpdg-pipi}, by {\nuovo 4.3 standard deviations.
More in detail, the BR of Eq.~\eqref{eq:bg-pipi} has been obtained by using the pion form factor at the \jp\ mass
\be
|F_\pi(M_\jp^2)|_{\textsc{BaBar}}=0.057\pm 0.010\,.
\nen
This value has been extrapolated from the \textsc{BaBar} data, which cover the interval $0.305\,{\rm GeV}\le\sqrt{q^2}\le 2.950\,{\rm GeV}$, with the fit function and the parameters of Ref.~\cite{Lees:2012cj}. The error has been computed by propagating in quadrature the errors of the fit parameters with the standard procedure.
}
\\
The obtained value of $\bg(\pipi)$ unavoidably means that there must be a further contribution.
Since the purely strong three-gluon amplitude, \aggg, is suppressed by \gp\ conservation, the remaining amplitude that, contrary to what commonly expected, could play an important role is the one related to the second diagram of Fig.~\ref{fig:3.states.h}, i.e., \agg. Moreover, having two sizable amplitudes, there could also be a constructive interference term that would help in reconciling the prediction and the measured value for the $\jp\to\pipi$ BR.    
\\
The amplitude to be considered is then 
\be
\a(\pipi)=\ag(\pipi)+\agg(\pipi)\,,
\nen 
so that, following Eq.~\eqref{eq:br-contributions}, the prediction for the BR is
\be
\b(\pipi)=\bg(\pipi)+\bgg(\pipi)+\mathcal{I}(\pipi)\,.\,\,\,\,
\label{eq:bpi-2amp}
\en
The calculation of the amplitude $\agg(\pipi)$ in the framework of QCD is quite difficult because the hadronization of the two-gluon plus one-photon intermediate state into \pipi\ occurs at the few-GeV energy regime where QCD is still not perturbative. 
\\
In this paper we calculate the imaginary part of the amplitude $\agg(\pipi)$ due to the dominant intermediate states, by means of a phenomenological procedure, based on the Cutkosky rule~\cite{cut} and experimental rates of the involved \jp\ decays.
Using such an amplitude we obtain a lower limit for $\bgg(\pipi)$ and show that, within the errors, it is of the same order of $\bg(\pipi)$.
\section{The decay channel $J/\psi \to \pi^+ \pi^-$}
\label{sec:th.decay.2pi}
As already pointed out, the total BR for the \gp-violating decay $\jp\to\pipi$ can be parametrized as given in Eq.~\eqref{eq:bpi-2amp}, where, besides the dominant one-photon contribution $\bg(\pipi)$, also $\bgg(\pipi)$ is taken into account. 
\\
In terms of amplitudes we can write
\be
|\a|^2 = |\ag|^2 + |\agg|^2 + 2|\ag| |\agg| \cos (\varphi) \, ,
\label{eq:a^2}
\en
where
\be
\varphi=\arg(\agg)-\arg(\ag)=\phi_{gg\gamma}-\phi_\gamma\,,
\label{eq:rel-phase}
\en 
is the relative phase between the two amplitudes, being $\phi_{X}$ the absolute phase of the amplitude $\mathcal A_X$.
In Eq.~\eqref{eq:a^2} and in the following the symbol $(\pipi)$ is omitted, being understood that all amplitudes and BRs refer to the \pipi\ channel.
\\
The BR is obtained as
{\nuovo\be
\b &=& {1 \over 2M_{J/\psi} \Gamma_{J/\psi}} \!\!\int\!\! d\rho_2 \, \overline{|\mathcal A|^2} = {\sqrt{M^2_{J/\psi}\!-\!4M_{\pi}^2} \over 64 \pi^2 M^2_{J/\psi} \Gamma_{J/\psi}} \!\!\int\!\! d\Omega \, \overline{|\mathcal A|^2}  \nonumber \\
&=& \b_{gg \gamma} \!+\! \b_{\gamma} \!+\! \underbrace{{\cos (\varphi) \sqrt{M^2_{J/\psi}\!-\!4M_{\pi}^2} \over 32 \pi^2 M^2_{J/\psi} \Gamma_{J/\psi}}\!\! \int\!\! d\Omega \, \overline{|\mathcal A_{gg \gamma}| |\mathcal A_{\gamma}|} }_{\mbox{interference term }\mathcal{I}} \, ,
\nen}%
where: $d\rho_2$ is the element of the two-body phase space, $M_\pi$ is the charged pion mass, $M_\jp$ and $\Gamma_\jp$ are mass and width of \jp.
\\
Moreover, to highlight the contribution due to the real part and that to the imaginary part of \agg, which is computed in the next section, the BR \bgg\ can be decomposed as
\be
\bgg &=& \bgg^{\re} + \bgg^{\im} = {1 \over 2M_{J/\psi} \Gamma_{J/\psi}} \label{eq:b-gg}\\
&\times &  \left(\int \!\!d\rho_2 \, \overline{\big(\re(\agg)\big)^2} \!+ \!\!\int\!\! d\rho_2 \, \overline{\big(\im(\agg)\big)^2} \right) \, .
\nen
\section{The contribution $\im(\agg)$}
\label{sec:th.calc}
\begin{figure}[h]
\begin{center}
	\includegraphics[width=60 mm]{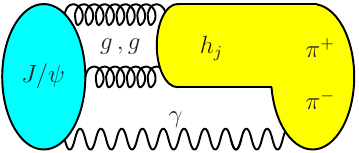}	
\caption{\label{fig:ozi2} Two-gluon plus one-photon mechanism of the decay $\jp\to\pipi$ with light unflavored mesons $h_j$ in the intermediate states. The light blue and the yellow areas indicate the domains of the charm and light quarks, respectively.}
\end{center}	
\end{figure}

\begin{figure}[h]
\begin{center}
	\includegraphics[width=60 mm]{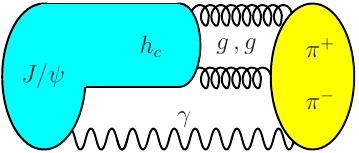}	
\caption{\label{fig:ozi3} Decay mechanism of the \jp\ into \pipi\ mediated by strongly coupled charmonia $h_c$. The color scheme is the same of Fig.~\ref{fig:ozi2}.}
\end{center}	
\end{figure}

\begin{figure}[h]
\begin{center}
	\includegraphics[width=60 mm]{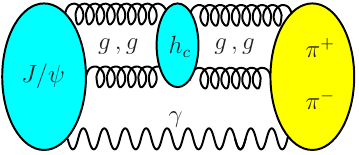}	
\caption{\label{fig:ozi1} OZI double-suppressed \jp\ decay mechanism with charmonia $h_c$ in the intermediate states. The color scheme is the same of Fig.~\ref{fig:ozi2}.}
\end{center}	
\end{figure}

\begin{figure}[h]
\begin{center}
	\includegraphics[width=60 mm]{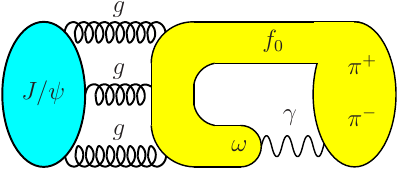}	
\caption{\label{fig:ozi3g} \hl{Decay mechanism for $\jp\to f_0(980)\omega\to f_0(980)\gamma\to\pipi$. The light blue and the yellow areas highlight the domains of the charm and light quarks, respectively. The double vertical lines indicate the cut that separates the full decay into the two sub-processes, used to apply the Cutkosky rule.}}
\end{center}	
\end{figure}

In order to use the Cutkosky rule~\cite{cut} to calculate the imaginary part of the amplitude $\agg$, it needs to consider all possible on-shell intermediate states that can contribute to the decay chain $\jp\to (gg \gamma)^* \to \pipi$. We consider the {\nuovo mechanism where the two gluons hadronize into a set  $\mathcal{P}$ of \hl{$C=+1$} mesons $h_j$}, so that the decay proceeds as
{\nuovo\be
\jp \to \sum_{h_j\in \mathcal{P}} (h_j \gamma)^* \to \pi^+ \pi^- \,.
\nen}%
\hl{(See Ref.~\cite{Field:2001iu} for a detailed analysis of radiative decays $\jp\to\gamma+$\,hadrons).}
The elements $h_j$ of the set $\mathcal{P}$ are only light unflavored mesons, that then couple strongly (OZI-allowed process) with the \pipi\ final state. Indeed, only in these cases, the underlying mechanism, sketched in Fig.~\ref{fig:ozi2}, is the one we want to evaluate. More in detail, such a process consists in the sequence of two conversions: the OZI suppressed coupling of the  \jp\ to $h_j$, via two gluons, and the OZI allowed \pipi\ production mediated by $h_j$ and the "spectator" photon. 
\\
\hl{Intermediate states with charmonia are excluded because they proceed through a different mechanism. For instance, the case with $h_j=\eta_c$, shown in Fig.~\ref{fig:ozi3}, is characterized by a first radiative conversion of the \jp\ into $\eta_c$, followed by the OZI-suppressed coupling to the \pipi\ final state. This means that we are evaluating the $gg\gamma$ coupling of the $\eta_c\gamma$ to the \pipi, rather than that of the $\jp\gamma$.}
\\
Another possible class of decay mechanisms, passing through the ``wanted'' contribution, is the one sketched in Fig.~\ref{fig:ozi1}. Apart from the computational difficulty, this contribution is double-OZI-suppressed so it is negligible with respect to that of Fig.~\ref{fig:ozi2}.    
\\
\hl{Following the same argument,  non-radiative, light-quark intermediate states, as, e.g., $f_0(980)\omega$, that subsequently scatters into $f_0(980)\gamma$, via $\omega\to\gamma$ conversion, can not be used because they proceed dominantly through the three-gluon channel, as sketched in Fig.~\ref{fig:ozi3g}.} 
\\
Using the Cutkosky rule~\cite{cut} the imaginary part of \agg\ is given in terms of a series on the intermediate states $h_j\gamma$, i.e.,
\be
\im(\agg) = \frac{1}{2} \!\sum_j \SumInt\! d\rho \,  \a^*\!(\jp \!\to\! h_j \gamma) \a(\pipi\!\!\! \to\! h_j \gamma) \, ,
\no\\
\label{eq:cut0}
\en
where the internal-sum runs over the photon polarizations and the integration is on the phase space
\be
\label{eq:PS}
d\rho = \frac{p^0}{4 \pi M_{J/\psi}} \, \frac{d\Omega}{4 \pi} \, ,
\en
being $p^\mu$ the four-momentum of the photon.
\subsection{\hl{Selection of intermediate mesons}}
\begin{figure}[h]
\begin{center}
	\includegraphics[width=\columnwidth]{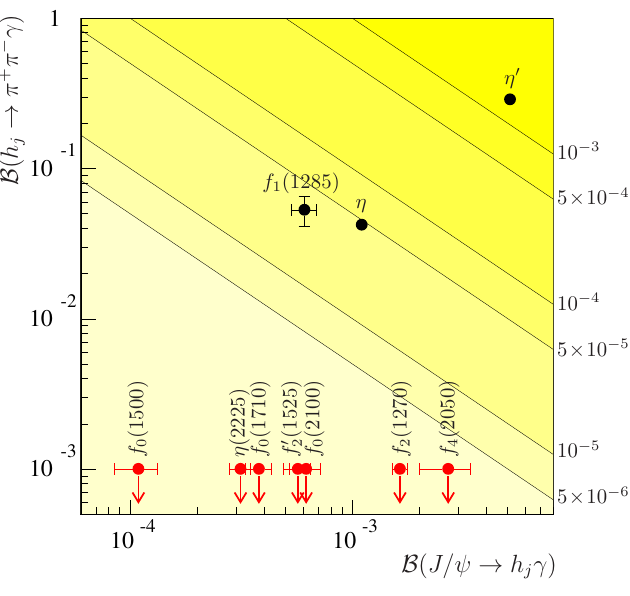}	
\caption{\hl{\label{fig:branchings} Combined BRs of most probable intermediate states. 	The black points represent the three mesons for which data on both branching fractions are available. Points for all the other mesons are aligned along the line $\b(\pipi\to h_j\gamma)=10^{-3}$ in order to make them visible in logarithmic scale. Oblique lines are hyperbola, i.e., geometric loci of all points having the  product $\b(\jp\to h_j\gamma)\cdot\b(\pipi\to h_j\gamma)$ constantly equal to the value reported on the right side.}}
\end{center}	
\end{figure}

\hl{The selection of all the possible intermediate channel, i.e., of all possible meson $h_j$ with $C=+1$, is experimentally driven. Table~\ref{tab:1} reports all the branching ratios listed in Ref.~\cite{pdg}. While there are ten candidates on the \jp\ side, only three sets of data are available on the \pipi\ side. As a first estimate of the contribution that each channel can give one could consider the product of the BRs, i.e., $\b(\jp\to h_j\gamma)\cdot\b(\pipi\to h_j\gamma)$. In Fig.~\ref{fig:branchings} all meson $h_j$ have been mapped in the 
	$\b(\jp\to h_j\gamma)-\b(\pipi\to h_j\gamma)$ plane. Mesons bringing higher contribution lie on the upper right corner and are represented by black solid circles.}
\hl{The most prominent contribution, well above the hyperbola at $10^{-3}$, see Fig.~\ref{fig:branchings}, is the one due to the $\eta'$ meson, that couples strongly to both, \jp\ and \pipi. The $\eta$ meson contribution lies well below, around the hyperbola at $5\times 10^{-5}$, having $\b(\jp\to \eta\gamma)\cdot\b(\pipi\to \eta\gamma)\simeq 4.66\times 10^{-5} $.  
		\\
		A further contribution that could be considered is the one due to the axial vector meson \f, for which the combined strength is compatible with that of the $\eta$ meson, indeed:
		$\b(\jp\to f_1)\cdot\b(\pipi\to f_1\gamma)\simeq 3.23\times 10^{-5}$.}
	\begin{table}[h]
	\begin{center}
			\begin{tabular}{r|r|r|r}
 Meson $M$ & $J^{PC}$ & $10^3\cdot\b(\jp\to h_j\gamma)$ & $10^3\cdot\b(h_j\to\pipi\gamma)$ \\
\hline\hline
$\eta$ & $0^{-+}$  & $1.104\pm 0.034$ & $42.2\pm 0.8$\\
\hline
$\eta'(958)$ & $0^{++}$ &  $5.13\pm 0.17$ & $289\pm 50$\\
\hline
$f_2(1270)$ & $2^{++}$  & $1.64\pm 0.12$ & no data\\
\hline
$f_1(1285)$ & $1^{++}$  & $0.61\pm 0.08$ & ($\rho^0$) $53\pm 12$\\
\hline
$f_0(1500)$ & $0^{++}$  & $0.109\pm 0.024$ & no data\\
\hline
$f'_2(1525)$ & $2^{++}$  & $0.57^{+0.08}_{-0.05}$ & no data\\
\hline
$f_0(1710)$ & $0^{++}$  & $0.38\pm 0.05$ & no data\\
\hline
$f_4(2050)$ & $4^{++}$ & $2.7\pm 0.7$ & no data\\
\hline
$f_0(2100)$ & $0^{++}$  & $0.62\pm 0.10$ & no data\\
\hline
$\eta(2225)$ & $0^{-+}$  & $0.314^{+0.050}_{-0.019}$ & no data\\
\hline
\hline
\end{tabular}
\caption{\label{tab:1}Branching ratios of a selection of intermediate decays~\cite{pdg}.}\end{center}
\end{table}
\\
In light of that, the imaginary part of the amplitude $\agg$ has  \hl{three main} contributions, i.e.,
\hl{\be
\im(\agg)&\simeq&
\frac{1}{2} \!\SumInt\! d\rho \,  \a^*\!(\jp \!\to\! \eta\gamma) \a(\pipi\!\!\! \to\! \eta\gamma) \no\\
&&+
\frac{1}{2} \!\SumInt\! d\rho \,  \a^*\!(\jp \!\to\! \epg) \a(\pipi\!\!\! \to\! \epg) \no\\
&&+
\frac{1}{2} \!\SumInt\! d\rho \,  \a^*\!(\jp \!\to\! f_1\gamma) \a(\pipi\!\!\! \to\! f_1\gamma)\no\\
\im(\agg)&\simeq&
\im(\a_{\eta'\gamma})+\im(\a_{\eta\gamma})+\im(\a_{f_1\gamma})\,,
\label{eq:ImAgg}
\en}%
\hl{where $f_1$ here and in the following stands for the \f\ meson and the approximate identity is due to the truncation of the series.}
%
%
%
\subsection{A phenomenological calculation based on the Cutkosky rule}
\label{subsec:calculation}
The first amplitude in the right-hand-side of Eq.~\eqref{eq:cut0}
is that of the decay
\be
\jp(P) \to h_j(k) + \gamma(p)\,,
\nen
where, in parentheses, are reported the four-momenta, \hl{and $h_j$ can be either a pseudoscalar, $\eta$ and $\eta'$, or the axial vector meson $f_1$}. By invoking gauge and Lorentz invariance, the \hl{amplitudes of the radiative decays of a vector meson into a pseudoscalar, $\eta$, and into an axial vector meson, $f_1$, can be written as~\cite{Pacetti:2009pg,Rudenko:2017bel}
\be
\begin{array}{rcl}
\mathcal A(\jp \to \eg) &=& g_{\eg}^{\jp}  p_{\tau} P_{\lambda} \epsilon_\delta (\jp) \epsilon_\sigma (\gamma) \varepsilon^{\tau \lambda \delta \sigma} \, ,\\
&&\\
\mathcal A(\jp \to \fg) &=& g_{\fg}^{\jp} p_{\tau} \epsilon_\lambda(f_1) \epsilon_\delta (\jp) \epsilon_\sigma (\gamma) \varepsilon^{\tau \lambda \delta \sigma} \, ,\\
\end{array}
\label{eq:vertexPV}
\en
where $g_{\eg}^{J/\psi}$ and $g_{\fg}^{J/\psi}$ are the coupling constants, $\epsilon_\delta (\jp)$, $\epsilon_\sigma (\gamma)$ and $\epsilon_\lambda (f_1)$ are the \jp, photon and axial vector polarization vectors, and $\varepsilon^{\tau \lambda \delta \sigma}$ is the Levi-Civita symbol.}
\begin{figure}[h!]
\begin{center}
	  \includegraphics[width=70 mm]{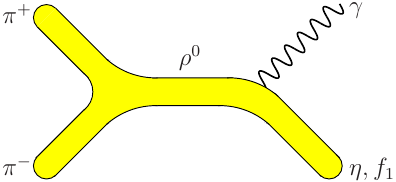}
\end{center}
\caption{Feynman diagram for $\pi^+ \pi^- \to \eg$ and $\pi^+ \pi^- \to \fg$ mediated by the $\rho^0$ meson.}
\label{fig:pi.to.eta.gamma}       
\end{figure}\\
The second amplitude in the right-hand-side of Eq.~\eqref{eq:cut0} concerns the \pipi\ annihilation process
\be
\pi^+(k_1)+\pi^-(k_2) \to h_j(k) +\gamma(p)\,.
\label{eq:pipi-geta}
\en
The amplitude for this process can be computed in terms of effective meson fields, as described by the Feynman diagram of Fig.~\ref{fig:pi.to.eta.gamma}. Here the coupling between the \pipi\ initial state and the $h_j\gamma$ final state is assumed to be mediated by the $\rho^0$ vector meson. Such an assumption is supported by the strong affinity of the two-pion system with quantum numbers $J^{PC}=1^{--}$ and the $\rho^0$, experimentally confirmed by the BR $\b(\rho^0 \to\pipi)= 1$~\cite{pdg}. \hl{It follows that the amplitudes read~\cite{Pacetti:2009pg,Rudenko:2017bel}
\be
\begin{array}{rcl}
\a(\pi\pi \to \eg) &\!\!=\!\!& \ds g_{\eg}^{\pi\pi} {d_{\alpha} p_{\beta}  \epsilon_\mu (\gamma) k_\nu \varepsilon^{ \alpha\beta \mu \nu} \over M_{\rho}^2 - q^2-iM_{\rho} \Gamma_{\rho}} \, ,\\
&&\\
\a(\pi\pi \to \fg) &\!\!=\!\!& \ds
 g_{\fg}^{\pi\pi} 
{d_{\alpha} p_{\beta}  \epsilon_\mu (\gamma)\epsilon_\nu(f_1) \varepsilon^{\alpha\beta \mu \nu} \over M_{\rho}^2 - q^2-iM_{\rho} \Gamma_{\rho}} \, ,\\
\end{array}
\label{eq:vertex-pp-pg}
\en
where $g^{\pipi}_{\eta(f_1)\gamma}$ is the $\pipi$-$\eta(f_1) \gamma$ coupling constant, $d=k_1-k_2$, while $q= k_1+k_2$, $M_\rho$ and $\Gamma_\rho$ are the four-momentum, the mass and the width of the $\rho^0$ meson.} In the following the imaginary term at denominator, $iM_{\rho} \Gamma_{\rho}$, will be omitted, because its contribution to the resulting BR is of the order of $0.01\%$ and then negligible with respect to the experimental uncertainty, $\sim 6\%$ (see Eq.~\eqref{eq:bgg-value}). Moreover, the negligibility of this term allows to recover the reality of $\im(\agg)$ \hl{by also validating the truncation of the Cutkosky series}.
\\
Using the amplitudes of Eqs.~\eqref{eq:vertexPV} and~\eqref{eq:vertex-pp-pg}, we compute the polarization sum of the Cutkosky formula of Eq.~\eqref{eq:cut0}
\be
\mathcal{Z}(h_j) \equiv \sum_{\rm pol} \a(\pipi \to h_j) \a^*(\jp \to h_j) \, ,
\nen
in the \jp\ and \pipi\ center of mass frame (CM), i.e., 
whit the four-momenta
\be 
\begin{array}{rcl}
P &\!\!=\!\!&  q= (M_{J/\psi},0,0,0)\,,\vspace{2mm}\\ 
p&\!\!=\!\!&(p^0,\vec p)=p^0(1, \sin (\theta), 0, \cos (\theta))\,,\vspace{2mm}\\
 k&\!\!=\!\!&(k^0,-\vec p)=(k^0, p^0\sin (\theta), 0, p^0\cos (\theta))\,,\vspace{2mm}\\
 k_{1,2} &\!\!=\!\!& (M_{J/\psi}/2,0,0, \pm\omega)\,,\\
\end{array}
\nen%
\hl{where $\theta$ is the scattering angle of the photon and $\omega$ is the modulus of the pion three-momenta.}
\\
\hl{The results for $\mathcal{Z}(\eta)$ and $\mathcal{Z}(f_1)$ are
\be
\mathcal Z (\eta)&=& g_{\eg}^{\pi\pi} g_{\eg}^{\jp}
\frac{M_{\jp}^2 - M_{\eta}^2}{M_{\jp}^2 -M_{\rho}^2}  
\lq \rule{0cm}{6mm}\omega\, M_{J/\psi}\cos (\theta)  \, p^\mu \right.\nonumber \\
&& \left.- \frac{M_{\jp}^2-M_\eta^2}{4}\, d^\mu \rq \epsilon_\mu(\jp)\,,
\no\\
\mathcal Z (f_1)&=& g_{\fg}^{\pi \pi} g_{\fg}^{\jp}
\frac{M_{\jp}^2 - M_{f_1}^2}{M_{\jp}^2-M_{\rho}^2} \no\\
&&\times
\lq 
\frac{\omega\cos(\theta)}{M_{\jp}} 
\lt\frac{M_{\jp}^2}{M_{f_1}^2}-2\rt
 p^\mu 
 \right.\nonumber \\
&& - \frac{1}{4}\lt \frac{M_{\jp}^2}{M_{f_1}^2}-1\rt
d^\mu \Bigg]\epsilon_\mu (J/\psi) \,,
\nen}
\hl{that, using Eqs.~\eqref{eq:PS} and~\eqref{eq:ImAgg}, give the imaginary parts
\be
\im(\a_{\eg}) &=& \frac{1}{2}\int d\rho\, \mathcal{Z}(\eta)
=\frac{p^0}{4 \pi M_{\jp}} \int \frac{d\Omega}{4 \pi}\,\mathcal{Z}(\eta)
\no\\
&=&
\sqrt{{M_{J/\psi}^2 \over 4}-M_{\pi}^2}\,
 \frac{g_{\eg}^{\pi\pi} g_{\eg}^{\jp} M_{\jp}^4 \epsilon_3(\jp)}{48\pi\lt M_{\jp}^2-M_{\rho}^2\rt} 
 \nonumber \\
&&\times  \lt 1-\frac{M_{\eta}^2}{M^2_{\jp}} \rt^{3}\,,
\no\\
&&\label{eq:im-etp}
\\
\no\\
\im(\a_{\fg}) &=& \frac{1}{2}\int d\rho\, \mathcal{Z}(f_1)
=\frac{p^0}{4 \pi M_{\jp}} \int \frac{d\Omega}{4 \pi}\,\mathcal{Z}(f_1)
\no\\
&=&
\sqrt{{M_{J/\psi}^2 \over 4}-M_{\pi}^2}\,
 \frac{g_{\fg}^{\pi\pi} g_{\fg}^{\jp} M_{\jp}^4 \epsilon_3(\jp)}{48\pi M_{f_1}^2\lt M_{\jp}^2-M_{\rho}^2\rt} 
 \nonumber \\
&&\times  \lt 1-\frac{M_{f_1}^2}{M^2_{\jp}} \rt^{3}
\lt 1+\frac{M_{f_1}^2}{M^2_{\jp}} \rt\,,
\nen
where $\epsilon_3(\jp)=\epsilon_3^{(\sigma)}(\jp)$ is the numerical third component ($\mu=3$) of the generic $\sigma$-th polarization four-vector of the $\jp$ meson. 
The imaginary parts of Eq.~\eqref{eq:im-etp} and that due to the \epg\ intermediate state, which has the same structure of $\im(\a_{\eg})$, have to be summed up to obtain the complete imaginary of \agg, see Eq.~\eqref{eq:ImAgg}. The corresponding contribution to the BR, $\bgg^{\im}$, as given in Eq.~\eqref{eq:b-gg}, is
\be
\mathcal B^{\rm Im}_{gg \gamma} &=& \frac{\sqrt{M^2_{J/\psi}-4M_{\pi}^2}}{16 \pi M^2_{J/\psi} \Gamma_{J/\psi}} \, \overline{|\im(\agg)|^2}
\label{eq:bim-gg}
\\
&=&\frac{\left(M^2_{\jp}-4M_{\pi}^2\right)^{3/2}}{4(48\pi)^3 M^6_{\jp} \Gamma_{J/\psi}} 
\frac{\left|
\ds\sum_{h=\eta,\eta',f_1} g_{h\gamma}^{\pi\pi} g_{h \gamma}^{\jp} K_h  \right|^{2}}{\left(M^2_{J/\psi} - M_{\rho}^2\right)^{2}}\,,
\nen
where the average over the polarization states of the $\jp$ meson has been performed and $K_h$ is the kinematical quantity
\be
K_h=\left\{
\begin{array}{ll}
\ds	\lt M_{\jp}^2-M_h^2\rt^3 & h=\eta,\eta'\\
\ds \frac{\lt M_{\jp}^2-M_h^2\rt^3}{M_h^2}\lt1+\frac{M_{h}^2}{M_{\jp}^2}\rt & h=f_1\\
\end{array}
\right..\,\,\,\,\,\,\,\,\,\,
\label{eq:Kh}
\en}%
The quantity of Eq.~\eqref{eq:bim-gg} represents a lower limit for \bgg, \hl{because the contribution due to the real part of the amplitude, $\bgg^{\re}$, as shown in Eq.~\eqref{eq:b-gg}, is positive.} 
\\ 
In order to obtain the numerical value of $\bgg^{\im}$, apart from the \jp\ width and the masses of all mesons involved that are well known, the values of the six coupling constants $g^{\pi\pi}_{h \gamma}$ and $g^{\jp}_{h\gamma}$ ($h=\eta$, $\eta', f_1$) have to be estimated experimentally.   
\subsection{\hl{The $g^\jp_{\eg}$, $g^\jp_{\epg}$ and $g^\jp_{\fg}$ coupling constants}}
\label{subsec:coup.cost.1}
\hl{The experimental value of the modulus of the coupling constant $g_{h\gamma}^{J/\psi}$, with $h=\eta$, $\eta'$, $f_1$, can be extracted from the rate of the corresponding radiative decay $\jp \to h \gamma$. Using the amplitudes of Eq.~\eqref{eq:vertexPV}, the radiative decay width is
\be
\Gamma (\jp \to h \gamma) &=& \frac{K_h}{96\pi M_{\jp}^3} 
 \big|g_{h\gamma}^{J/\psi}\big|^2 \, ,
\nen
where $K_h$ is the kinematical quantity defined in Eq.~\eqref{eq:Kh}. The modulus of the coupling constant can be extracted as
\be
\big|g_{h \gamma}^{J/\psi}\big| = \sqrt{\frac{96 \pi M_{\jp}^3 \Gamma (J/\psi \to h\gamma)}{K_h}}\,.
\nen}%
Finally, by using the experimental values of radiative decay widths $\Gamma(\jp\to\eta\gamma)$, $\Gamma(\jp\to\epg)$ and  
 $\Gamma(\jp\to\fg)$~\cite{pdg} of
that are also reported in the first two rows of tab.~\ref{tab:Gamma.pdg}, the coupling constants are
\be
|g_{h \gamma}^{J/\psi}|\! =\!
\left\{\!\begin{array}{ll}
(1.070 \!\pm\! 0.023) \!\times \!10^{-3} \ {\rm GeV^{-1}} &\hh h=\eta \\
&\\
(2.563 \!\pm\! 0.055) \!\times \!10^{-3} \ {\rm GeV^{-1}} & \hh h=\eta'\\
&\\
\hl{(1.191 \!\pm\! 0.080) \!\times \!10^{-3} }& \hh h=f_1\\
\end{array}\right. \!.\,\,\,\,\,\,\,\,\,
\label{eq:cc-eg}
\en
\hl{
Here and in the following we consider an error with two significant figures in light of further manipulations.
It is interesting to notice that while the coupling constant of the axial vector is adimensional, those of the pseudoscalar mesons have the dimension of inverse energy. This is a consequence of the structure of the corresponding amplitudes, given in Eq.~\eqref{eq:vertexPV}. Indeed, they differ only by the interchange of the \jp\ four-momentum $P_\lambda$ with the adimensional polarization vector of $f_1$.}
\subsection{\hl{The $g^{\pi\pi}_{\eg}$, $g^{\pi\pi}_{\epg}$ and $g^{\pi\pi}_{\fg}$ coupling constants}}
\label{subsec:coup.cost.2}
The coupling constant $g_{h\gamma}^{\pi\pi}$, with $h=\eta$, $\eta'$ and $f_1$, can not be directly measured, because there are no data on the cross section of the annihilation process $\pipi\to h\gamma$, whose pseudoscalar and axial vector amplitudes, defined in Eq.~\eqref{eq:vertex-pp-pg}, have been parametrized in terms of such coupling constants. 
Nevertheless, as a consequence of the crossing symmetry, the same coupling constants must appear in the amplitudes of the decay 
\be
h(k) \to \pi^+(k_1) + \pi^-(k_2) + \gamma(\tilde p)\,,\hh\hh
h=\eta,\eta', f_1\,.
\nen
This decay is obtained by moving the photon from the final to the initial state of the original reaction of Eq.~\eqref{eq:pipi-geta}, with the Feynman diagram of Fig.~\ref{fig:pi.to.eta.gamma}, and then by making a time-reversal transformation.
\\
\hl{Therefore, by using the amplitudes of Eq.~\eqref{eq:vertex-pp-pg}, the decay width is 
\be
\Gamma(h \to \pi^+ \pi^- \gamma) 
&=& 
\int 
 \overline{|\mathcal A(h \to \pi^+ \pi^- \gamma)|^2} \,d\rho_3
\no\\
&=& \frac{1}{(2\pi)^3} \frac{|g_{h \gamma}^{\pi \pi}|^2 }{128M_{h}^3}\label{eq:Gpipig}\\
&&\times\int_{q^2_{\rm min}}^{q^2_{\rm max}} \!\!\!\!\!dq^2 \!\!\!
\int_{{q_1^2}_{\rm min}(q^2)}^{{q_1^2}_{\rm max}(q^2)} \!\!\!\!\!dq_1^2\, I_{h}(q^2,q_1^2)\,,
\nen
where $d\rho_3$ is the three-body phase space and, the integration variables and corresponding limits are: $q^2 \equiv (k-\tilde p)^2=(k_1+k_2)^2$, $q^2_1 \equiv (k_1+\tilde p)^2=(k-k_2)^2$, 
\be
&
q^2_{\rm min} = 4M_{\pi}^2\,,\hh\hh q^2_{\rm max} = M_{h}^2\,,&\no\\
&
{q_1^2}_{\rm min,max}(q^2) =\ds {M_{h}^4 \over 4q^2} - \left( \sqrt{{q^2 \over 4}-M_{\pi}^2} \pm \frac{M_{h}^2-q^2}{2\sqrt{q^2}} \right)^2\,,&
\nen
with $h=\eta$, $\eta'$, $f_1$. The functions $I_h(q^2,q_1^2)$ have two different forms, in the case of pseudoscalar mesons 
\be  
I_h(q^2,q_1^2)
&=&\frac{(q^2\!-\!4M_{\pi}^2)\big(q^2\!-\!M_{h}^2\big)^2\!}{\big(q^2\!-\! M_{\rho}^2\big)^2 \!+\!\Gamma_{\rho}^2 M_{\rho}^2} \no \\
&&-\ds\frac{q^2\big(q^2\!+\! 2q_1^2 \!-\! 2M_{\pi}^2\! -\! M_{h}^2 \big)^2}{\big(q^2\!-\! M_{\rho}^2\big)^2 \!+\!\Gamma_{\rho}^2 M_{\rho}^2}\,,\hh h=\eta,\eta'\,,
\nen
while for the axial vector meson it reads
\be
I_{f_1}(q^2,q_1^2)
&=&\frac{1}{3M_{f_1}^2}\lq\frac{(q^2\!-\!4M_{\pi}^2)\big(q^2\!-\!M_{f_1}^2\big)^2\!}{\big(q^2\!-\! M_{\rho}^2\big)^2 \!+\!\Gamma_{\rho}^2 M_{\rho}^2} \right.  \no\\
&&\left.-\frac{\big( q^2-2M_{f_1}^2\big)\big(q^2\!+\! 2q_1^2 \!-\! 2M_{\pi}^2\! -\! M_{f_1}^2 \big)^2}{\big(q^2\!-\! M_{\rho}^2\big)^2 \!+\!\Gamma_{\rho}^2 M_{\rho}^2}\rq \,.
\nen}%
The phase-space integrals are
\be
\tilde I_h&=&\int_{q^2_{\rm min}}^{q^2_{\rm max}} \!\!\!\!\!dq^2 \!\!\!
\int_{{q_1^2}_{\rm min}(q^2)}^{{q_1^2}_{\rm max}(q^2)} \!\!\!\!\!dq_1^2\, I_{h}(q^2,q_1^2)\no\\
&=&\left\{\begin{array}{ll}	
	(5.840 \pm 0.011) \times 10^{-5} \, \rm GeV^6 &\hh\hh h=\eta\\
	&\\
	(2.719 \pm 0.019) \times 10^{-1}\,\rm GeV^6 &\hh\hh h=\eta'\\
	&\\
	\hl{(6.403 \pm 0.052) \,\rm GeV^4} &\hh\hh h=f_1\\
\end{array}
\right.\,,
\nen
\hl{and also in this case the contribution due to the axial vector meson has a different dimension, $E^4$ instead of $E^6$, as a consequence of the different structure of the amplitude, see Eq.~\eqref{eq:vertex-pp-pg}}. Finally, the corresponding coupling constants can be extracted by means of
\be
|g_{h\gamma}^{\pi\pi}| &=& (2\pi M_{h})^{3/2} \sqrt{\frac{128 \cdot \Gamma(h\to \pi^+ \pi^- \gamma)}{\tilde I_{h}}}\no\\
&=&\left\{\begin{array}{ll}	
	(2.223 \pm 0.047) \ {\rm GeV^{-1}} &\hh\hh h=\eta\\
	&\\
	(2.431 \pm 0.060) \ {\rm GeV^{-1}} &\hh\hh h=\eta'\\
	&\\
	\hl{3.55 \pm 0.41}  &\hh\hh h=f_1\\
\end{array}
\right.\,.
\label{eq:cc-pipi}
\en
%
%
\begin{table}
\centering
\caption{Decay widths~\cite{pdg} of the processes that have been used in our computation.}
\label{tab:Gamma.pdg} 
\begin{tabular}{ll} 
\hline\noalign{\smallskip}
Decay processes & Decay widths $\Gamma$ $({\rm GeV})$  \\
\noalign{\smallskip}\hline\noalign{\smallskip}
$J/\psi \to \eta \gamma$ & $(1.026 \pm 0.044) \cdot 10^{-7}$ \\
$J/\psi \to \eta' \gamma$ & $(4.78 \pm 0.14) \cdot 10^{-7}$ \\
\hl{$J/\psi \to f_1 \gamma$} & \hl{$(5.67 \pm 0.76) \cdot 10^{-8}$} \\
$\eta \to \pi^+ \pi^- \gamma$ & $(5.53 \pm 0.32) \cdot 10^{-8}$ \\
$\eta' \to \pi^+ \pi^- \gamma$ & $(5.76 \pm 0.10) \cdot 10^{-5}$ \\
\hl{$f_1 \to \pi^+ \pi^- \gamma$} & \hl{$(1.20 \pm 0.28) \cdot 10^{-4}$} \\
\noalign{\smallskip}\hline
\end{tabular}
\end{table}
\subsection{Calculation of $\bgg^{\im}(\pipi)$}
\label{subsec:coup.num}
To compute $\bgg^{\im}$ we use the expression of Eq.~\eqref{eq:bim-gg} \hl{which contains the sum of the three amplitudes due to the intermediate mesons $\eta$, $\eta'$ and $f_1$. In principle the coupling constants and hence the amplitudes are complex, then they can interfere.
The relative phase of the amplitudes of the two pseudoscalar contributions, being due to the $\eta$ meson and to its first excitation $\eta'$, is assumed to be zero, i.e., they add up with constructive interference. On the other hand, the relative phase between the axial vector amplitude and those of the pseudoscalar mesons cannot be inferred by phenomenological arguments.
\\
The single contributions can be obtained from Eq.~\eqref{eq:bim-gg} and are
\be
\bgg^{\im}(\eta) &=& (1.176 \pm 0.080) \times 10^{-6}\,,\no\\
\bgg^{\im}(\eta') &=& (5.34 \pm 0.38) \times 10^{-6}\,,\no\\
\bgg^{\im}(f_1) &=& (0.74 \pm 0.20) \times 10^{-6}\,.
\nen
They follow a hierarchy that reproduces the distribution based on BRs shown in Fig.~\ref{fig:branchings}. The total pseudoscalar contribution, assuming constructive interference, is
\be
\bgg^{\im}(\eta+\eta') &=& \lt\sqrt{\bgg^{\im}(\eta)}+\sqrt{\bgg^{\im}(\eta')}\rt^2
\no\\
&=& (1.152 \pm 0.066) \times 10^{-5}\,.
\label{eq:bgg-value}
\en
Concerning the $f_1$ contribution, the extreme cases 
of destructive and constructive interference give 
\be
\begin{array}{rcl}
\bgg^{\im}\lt \eta+\eta'- f_1\rt
&=&
(0.643\pm 0.074)\times 10^{-5}\,,\\
\bgg^{\im}\lt \eta+\eta'+ f_1\rt
&=&
(1.81\pm 0.12)\times 10^{-5}\,.\\
\end{array}
\label{eq:f1+-}
\en
}%
\hl{The fact that these values, which represent a lower limit for \bgg, lie between the 13\% and the 37\% of $\bg(\pipi) = (4.7 \pm 1.7) \times 10^{-5}$, see Eq.~\eqref{eq:bg-pipi}, leaves open the possibility that the total \bgg\ contribution would be of the same order of \bg.}
 \\
 Ultimately, using in Eq.~\eqref{eq:bpi-2amp} the \bgg\ decomposition of Eq.~\eqref{eq:b-gg}, the value of Eq.~\eqref{eq:bgg-value} for $\bgg^{\im}$, \hl{as an average of the two possibilities of Eq.\eqref{eq:f1+-}}, and the experimental datum for \bg, as given in Eq.~\eqref{eq:bg-pipi}, we get 
\be
\b(\pipi)&=&\bg(\pipi)+\bgg(\pipi)+\mathcal{I}(\pipi)\label{eq:Bth}\\
&=&(5.9 \pm 1.7) \!\times\! 10^{-5} \!+\!\bgg^{\re}(\pipi)\!+\!\mathcal{I}(\pipi)\,,
\nen
to be compared with the PDG datum~\cite{pdg}, given in Eq.~\eqref{eq:bpdg-pipi}, i.e.,
\be
\mathcal B_{\rm PDG}(\pi^+\pi^-) = (14.7 \pm 1.4) \times 10^{-5} \, .
\nen
{\nuovo
\subsection{The real part}
\label{subsec:real-part}
The procedure outlined in Sec.~\ref{sec:th.calc}, based on the Cutkosky rule given in Eq.~\eqref{eq:ImAgg} and on a suitable selection of the dominant intermediate states, allows to compute only the imaginary part of the amplitude \agg. By defining the $q^2$-dependent form of the imaginary part of this amplitude, $\im\lq \agg(q^2)\rq$, so that the obtained value $\im\lt\agg\rt\equiv\im\big[\agg(M_\jp^2)\big]$, and assuming analyticity, one can exploit dispersion relations (DRs) to compute the real part starting from the imaginary part. However, since the definition of $\im\big[\agg(q^2)\big]$ is model-dependent, the value of $\re\lt\agg\rt\equiv\re\big[\agg(M_\jp^2)\big]$, computed by means of DRs, will be affected by a large systematic error.\\
A possible calculation of $\re(\agg)$ \hl{due to the pseudoscalar contribution} is reported in App.~\ref{app:th.calc}.
} 
\section{Conclusions}
\label{concl}
The \gp-violating \jp\ decay into \pipi\ represents a useful testbed for what we called \bg-dominance hypothesis, which is summarized by the BR formula $\b(\pipi) \simeq \bg(\pipi)$, with (see Eq. \eqref{eq:b-gamma})
\be
\bg(\pipi)= \B(\mu^+ \mu^-) \, {\sigma(\ee \to \pipi) \over \sigma^0(\ee \to \mu^+ \mu^-)} \Bigg|_{q^2 = M_{\jp}^2}\!\!\!.\,\,\,\,\,\,\,\,
\label{eq:1g-dom}
\en
This follows from the assumption that the main contribution to the decay amplitude is that due to the one-photon exchange mechanism, the corresponding Feynman diagram is shown in the lower panel of Fig.~\ref{fig:3.states.h}. The other amplitudes contain gluons and hence are suppressed by \gp\ conservation.  
\\
The \bg-dominance is verified for other \gp-violating \jp\ decays, as those into four and six pions~\cite{Ferroli:2016zqf}. Indeed the BRs for these decays, computed by means of the formula of Eq.~\eqref{eq:b-gamma} using only cross section data, are in good agreement with the experimental rates~\cite{Ferroli:2016zqf}.
\\
However, for the simplest even-multipion final state, the \pipi\ one, a discrepancy of \hl{4.5} standard deviations is obtained between the PDG BR, given in Eq.~\eqref{eq:bpdg-pipi}, and the $\bg(\pipi)$ of Eq.~\eqref{eq:bg-pipi}.
\\
We considered the possibility that in this particular case the amplitude $\agg(\pipi)$, due to the presence of the photon, would not suffer the \gp\ suppression, being compatible with $\ag(\pipi)$, as expected in case of \gp-conserving decays.
\\
We have defined a phenomenological procedure, based on the Cutkosky rule, to compute the imaginary part of the amplitude \agg.
\hl{By considering the only $\eta\gamma$, $\eta'\gamma$ and $f_1\gamma$ intermediate states, that are phenomenologically the most probable ones (see Fig.~\ref{fig:branchings}), the contribution to the BR due to the $\im(\agg(\pipi))$ can vary between the values of Eq.~\eqref{eq:f1+-}. They are in average the 20\% of}
\be
\bg(\pipi)=(4.7\pm 1.7)\times 10^{-5}\,,
\nen
\hl{so that they could generate an interference effect of the same order of $\bg(\pipi)$, i.e.,}
\hl{\be
2\sqrt{\bg(\pipi)\bgg^\im(\eta+\eta')}=(4.6\pm0.8)\times 10^{-5}\,.
\nen
By considering also the $f_1$ contribution in the two extreme cases of Eq.~\eqref{eq:f1+-} the interference terms are 
\be
2\sqrt{\bg(\pipi)\bgg^\im(\eta+\eta'-f_1)}
&=&
(3.5\pm0.7)\times 10^{-5}\,,
\no\\
2\sqrt{\bg(\pipi)\bgg^\im(\eta+\eta'+f_1)}
&=&
(5.8\pm1.1)\times 10^{-5}\,.
\nen
These are large effects, indeed the BRs that can be obtained by composing with constructive interference $\bg(\pipi)$ and the only imaginary contribution $\bgg^\im(\eta+\eta'\pm f_1)$  are
\be
\b_\pm^{\im}
&=&\left\{
\begin{array}{l}
\lt\sqrt{\bg(\pipi)}\!+\!\sqrt{\bgg^\im(\eta+\eta'- f_1)}\rt^2\\
\lt\sqrt{\bg(\pipi)}\!+\!\sqrt{\bgg^\im(\eta+\eta'+ f_1)}\rt^2\\
\end{array}\right.
\no\\
&=&\left\{
\begin{array}{l}
	(8.8\pm2.3)\times 10^{-5} \\
	(12.3\pm2.8)\times 10^{-5}\\
\end{array}
\right.
\nen
}
which are compatible with the PDG datum of Eq.~\eqref{eq:bpdg-pipi}
\be
\b_{\rm PDG}(\pipi) = (14.7 \pm 1.4) \times 10^{-5}\,,
\nen
\hl{even though the real contribution $\bgg^\re(\eta+\eta'\pm f_1)$ has been not included.}
\\
In App.~\ref{app:th.calc} we proposed a procedure, based on DRs, to compute the real part of the amplitude \agg\  \hl{in the case of pseudoscalar intermediate mesons, i.e., by considering the functional form of the first expression of Eq.~\eqref{eq:im-etp}}. However, the result of such a computation depends on the $q^2$ functional form of the couplings, that has been defined on the basis of phenomenological arguments. Since this definition does not rely on first principles, the obtained result has a large systematic error that does not allow to draw any solid conclusion. 
\\
In summary, our main result is the definition of the procedure to compute the imaginary part of the amplitude \agg. It relies in the possibility of relating the $\jp\to h$ decay rate, $h$ stands for a hadronic state, to the rates of the "intermediate" radiative decays $\jp\to h_j\gamma$ and $h_j\to h\gamma$, where $h_j$ ($h_1$, $h_2$,\ldots) are \hl{$C=+1$} light mesons. In such a way, using as input the experimental BRs of the intermediate processes, the procedure automatically provides a phenomenological explanation for the validity of the $\bg$-dominance hypothesis. If, for a particular \gp-violating decay $\jp\to h$, there exists a set $\{h_j\}_{j=1}^n$ of \hl{$C=+1$} mesons with sizable rates $\Gamma(\jp\to h_j\gamma)$ and $\Gamma(h_j\to h\gamma)$, then the amplitude $\agg(h)$ will be of the same order of $\ag(h)$, i.e., there will be a violation of the $\bg(h)$-dominance rule.
\\
In the studied case with $h=\pipi$, we have considered \hl{three} mesons $h_1=\eta$, $h_2=\eta'$ and $h_3=f_1$, because the intermediate processes have large couplings, as indeed can be verified by considering the products of the corresponding BRs~\cite{pdg}, see Fig.~\ref{fig:branchings} and Table~\ref{tab:1},
\be
\begin{array}{rcl}
 \b(\jp\to\eta\gamma)\cdot\b(\eta\to\pipi\gamma)&\simeq& 5\times 10^{-5}\,,
\\
 \b(\jp\to\eta'\gamma)\cdot\b(\eta'\to\pipi\gamma)&\simeq& 1.5\times 10^{-3}\,,
\\
\hl{\b(\jp\to \fg)\cdot\b(f_1\to\pipi\gamma)}&\hl{\simeq}& \hl{3\times 10^{-5}}\,.\\
 \end{array}\,\,\,\,\,
\label{eq:brbr}
\en
\hl{On the contrary, in other cases as for instance the one with $h=2(\pipi)$, it appears quite evident that the contribution $\bgg^\im(2(\pipi))$ will be} suppressed having\footnote{\hl{Since there no data on these decay widths we have used the upper limit $\b(\eta'\to 2(\pipi)\,{\rm neutrals})<1\%$, in the first case, and $\b(f_1\to2(\pipi)\gamma)=\b(f_1\to\eta \pipi)\b(\eta\to\pipi\gamma)$~\cite{pdg}, in the second case.}}
\hl{\be
 \b(\jp\to\eta'\gamma)\cdot\b(\eta'\to2(\pipi)\gamma)&<& 5\times 10^{-5}\,,
 \\
\hl{\b(\jp\to \fg)\cdot\b(f_1\to2(\pipi)\gamma)}&\hl{\simeq}& \hl{10^{-5}}\,.
\nen
Even though there are no data on $\b(\eta\to 2(\pipi)\gamma)$, indeed the radiative decay $\eta\to2(\pipi)\gamma$ kinematically forbidden, in the light of the two-pion results for the coupling constants reported in Table.~\ref{tab:Gamma.pdg}, we expect that $|g^{4\pi}_{\eta\gamma}|<|g^{4\pi}_{\eta'\gamma}|$. It follows that $\agg(2(\pipi))$ is suppressed with respect to the corresponding two-pion amplitude $\agg(\pipi)$.}
\\
In principle the proposed procedure could be used to compute the amplitude \agg\ of all quarkonium decays in which \gp\ is violated. \hl{However, at higher masses, as those of bottomonia, the contribution  \bgg\ becomes almost negligible. Indeed $\bgg\simeq\bg$ $\propto \sigma(\ee\to\pipi)$, and the cross section at the bottomonium masses is very tiny because, as the mass $M$ diverges, it vanishes like $1/M^6$.}   
%
%
%
%
\appendix
{\nuovo%
\section{The one-photon decay rate}
\label{app:master-formula}
The formula reported in Eq.~\eqref{eq:b-gamma} gives exactly the one-photon decay rate of the \jp\ into a hadronic final state $h$. It has been used also elsewhere, e.g., in Eq.~(1) of Ref.~\cite{Milana:1993wk}. The simple expression can be explicitly obtained by considering, for instance, the \pipi\ hadronic final state. 
The one-photon mediated decay $\jp\to\pipi$, described by the Feynman diagram of Fig.~\ref{fig:1}, has rate
\be
\Gamma_\gamma(\pipi)=\frac{\alpha |G_\psi|^2}{12 M_{\jp}^3}
\left(\!1\!-\!\frac{4M_\pi^2}{M_{\jp}^2}\!\right)^{3/2}\!\!\!\!\!|F_\pi(M_{\jp}^2)|^2\,,\,\,\,\,\,\,\,\,
\label{eq:rate-app}
\en
where $G_\psi$ is the $\jp-\gamma^*$ coupling and $F_\pi$ the pion form factor, symbolically represented, in Fig.~\ref{fig:1}, by a solid disc and a grey hexagon respectively.  %
\begin{figure}[h]
\begin{center}
	\includegraphics[height=25 mm]{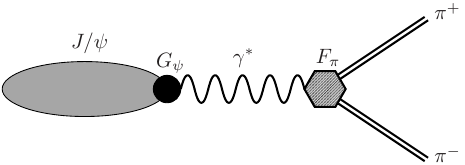}
\caption{\label{fig:1}Feynman diagram of the one-photon decay $\jp\to\pipi$. The solid disc and the grey hexagon represent the $\jp-\gamma^*$ coupling and the pion form factor.}
	\end{center}	
\end{figure}\\
The same coupling $G_\psi$ describes the decay $\jp\to\mu^+\mu^-$, which is purely electromagnetic. 
\begin{figure}[h]
\begin{center}
	\includegraphics[height=25 mm]{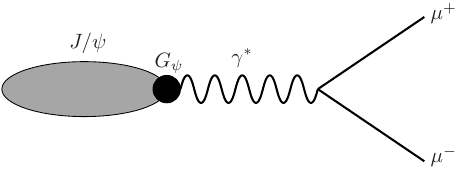}
	\caption{\label{fig:2}Feynman diagram of the purely electromagnetic decay $\jp\to\mu^+\mu^-$. The solid disc represents the $\jp-\gamma^*$ coupling.}
	\end{center}	
\end{figure}
\\
The Feynman diagram is shown in Fig.~\ref{fig:2} and the rate, neglecting the muon mass ($m_\mu\ll M_{\jp}$), is
\be
\Gamma(\mu^+\mu^-)=\frac{\alpha |G_\psi|^2}{3 M_{\jp}^3}
\,.
\nen
It depends on the value of the pion form factor at the \jp\ mass. Such a value, as well as all values of the pion form factor, have to be extracted from the Born dressed cross section of the annihilation process $\ee\to\pipi$, whose Feynman diagram is shown in Fig.~\ref{fig:3}.
The expression of the annihilation cross section is 
\be
\sigma_{\pipi}(q^2)=\frac{\pi\alpha^2}{3q^2}\left(1-\frac{4M_\pi^2}{q^2}\right)^{\!\!3/2}\!\!\!|F_\pi(q^2)|^2
\label{eq:rpipi}
\en
and it can be also written in terms of the $\ee\to\mu^+\mu^-$ bare cross section
\be
\sigma^0_{\mu^+\mu^-}(q^2)=\frac{4\pi\alpha^2}{3q^2}\,,
\nen
 as
\be
\sigma_{\pipi}(q^2)=\frac{\sigma^0_{\mu^+\mu^-}(q^2)}{4}
\left(1-\frac{4M_\pi^2}{q^2}\right)^{\!\!3/2}\!\!\!|F_\pi(q^2)|^2\,,\,\,\,\,\,\,\,\,
\label{eq:spipi}
\en
where the electron mass has been neglected ($m_e \ll m_\mu$)
and, for economy of symbols: $\sigma_f^{(0)}\equiv \sigma^{(0)}(\ee\to f)$, with $f=\pipi$, $\mu^+\mu^-$.
\begin{figure}[h]
\begin{center}
	\includegraphics[height=25 mm]{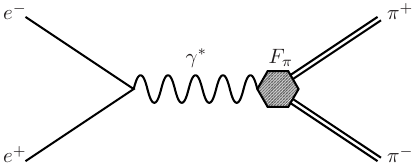}	
\caption{\label{fig:3}Feynman diagram of the annihilation $\ee\to\pipi$, in Born approximation. \hl{The the grey hexagon represents the pion form factor.}}
\end{center}	
\end{figure}\\
 It is important to stress that the pion form factor is extracted from the dressed cross section (e.g., Eq.~(24) of Ref.~\cite{Lees:2012cj}), hence, Eq.~\eqref{eq:rpipi} represents the $\ee\to\pipi$ cross section not corrected for the vacuum polarization contributions, called, indeed, dressed cross section.
\\
Finally, we use the cross section of Eq.~\eqref{eq:spipi} and 
the rate of Eq.~\eqref{eq:rpipi} to substitute the pion form factor times the velocity cube and modulus squared of the coupling divided by the \jp\ mass cube, respectively, in the \jp\ decay rate of Eq.~\eqref{eq:rate-app}. The obtained expression reads
\be
\Gamma_\gamma(\pipi)=\underbrace{\frac{\alpha |G_\psi|^2}{12 M_{\jp}^3}}_{\ds\frac{\Gamma(\mu^+\mu^-)}{4}}
\underbrace{\left(\!1\!-\!\frac{4M_\pi^2}{M_{\jp}^2}\!\right)^{3/2}\!\!\!\!\!|F_\pi(M_{\jp}^2)|^2}_{\ds\frac{4\sigma_{\pipi}(M_{\jp}^2)}{\sigma^0_{\mu^+\mu^-}(M_{\jp}^2)}}\,,
\nen
\hl{that, divided by the \jp\ total width, represents the formula of Eq.~\eqref{eq:b-gamma} in the case $h=\pipi$}, i.e.,
\be
\bg(\pipi)=\b(\mu^+\mu^-)\frac{\sigma_{\pipi}(M_\jp^2)}{\sigma^0_{\mu^+\mu^-}(M_\jp^2)}\,.
\nen

\section{A possible strategy to calculate $\bgg^{\re}(\pipi)$ \hl{in the pseudoscalar case}}
\label{app:th.calc}
The Cutkosky procedure, defined in Eq.~\eqref{eq:ImAgg} starting from the general form of Eq.~\eqref{eq:cut0}, allows to compute only the imaginary part of the amplitude \agg. 
\\
However, assuming analyticity for the amplitudes, DRs could be exploited to compute the real part of \agg, using as input its imaginary part as a function of $q^2$, i.e., the squared virtual mass of the \jp\ meson.
\\
Dispersion relations represent an analytic continuation procedure which is based on an integral representation. In more detail: given a function $f(z)$, analytic in the $z$ complex plane with the discontinuity cut $(x_0,\infty)$ over the positive real axis ($x_0>0$) and having the following properties:
\be
\begin{array}{lcl}
	f(x)\in\R &\hh&\forall x\in(-\infty,x_0)\,,\\
	&&\\ 
f(z)\ds\mathop{\propto}_{z\to x_0} (z-x_0)^\beta && \mbox{with }\re(\beta)>-1\,,\\
&&\\
f(z)=o\big(1/\ln(z)\big) && \mbox{as: }z\to\infty\,,\\
\end{array}
\label{eq:DR-cond}
\en 
then, $\forall x\in(x_0,\infty)$,
\be
\re[f(x)]=\frac{1}{\pi}\Pr\!\!\int_{x_0}^\infty\frac{\im[f(x')]}{x'-x}dx'\,,
\label{eq:DR-re}
\en
where the symbol $\Pr\!\!\int$ indicates the principal value integral, while $\re[f(x)]$ and $\im[f(x)]$ are real and imaginary values of the function on the upper edge of the cut.
\\
Assuming that the amplitude $\agg(q^2)$, as a function of $q^2$, fulfills the conditions of Eq.~\eqref{eq:DR-cond}, and using as lower threshold of the cut the value $q^2=4M_\pi^2$, 
the real part can be computed using Eq.~\eqref{eq:DR-re}, i.e.,
\be
\re[\agg(M_{\jp}^2)] =  \frac{1}{\pi}\Pr\!\! \int_{4 M_\pi^2}^\infty \frac{\im[ \agg(q^2)]}{q^2-M_{J/\psi}^2}\,dq^2  \,.\,\,\,\,\,\,\,\,
\label{eq:DR-re1}
\en
The imaginary part of \agg\ can be written using the decomposition of Eq.~\eqref{eq:ImAgg} and the terms, defined for $\eta\gamma$ and \epg\ channels following \hl{the first expression of} Eq.~\eqref{eq:im-etp}, with some change to account for the required $q^2$ dependence,
\be
  \im(\agg^{\eta+\eta'})&=&{\epsilon_3^\jp\sqrt{M_{J/\psi}^2 - 4M_{\pi}^2} \over 96 \pi M_{J/\psi}^2} \no \\
&&\times\frac{\ds\sum_{h=\eta,\eta'}g_{h\gamma}^{\pi \pi} g_{h \gamma}^{J/\psi} \! \left( M^2_{J/\psi}-M_{h}^2 \right)^{3}}{M_\jp^2-M_\rho^2} \, .
\label{eq:im-agg-cost}
\en
The procedure to determine the $q^2$-dependent form of $\im(\agg^{\eta+\eta'})$ consists not only in making the substitution $M_\jp^2\to q^2$, but also in introducing $q^2$-dependent couplings.
\\
Concerning $g^{\pi\pi}_{\eta\gamma}$ and $g^{\pi\pi}_{\epg}$, the dependence on $q^2$ is given by the $\rho^0$ propagator, so that the substitutions are
\be
g_{h\gamma}^{\pi\pi}\hh\to\hh g_{h\gamma}^{\pi\pi}\frac{|D_\rho(M_\jp^2)|}{|D_\rho(q^2)|}\,,\hh\hh h=\eta,\eta'\,,
\label{eq:pipi-q2}
\en
where $D_\rho(q^2)$ is the inverse Flatt\'e propagator~\cite{flatte} of the $\rho^0$, defined as
\be
D_\rho(q^2)=q^2 - M_{\rho}^2 + i  M_\rho \Gamma_{\rho}\bigg( {q^2-4 M_\pi^2 \over M_{\rho}^2-4 M_\pi^2} \bigg)^{3/2} \,,
\label{eq:flatte}
\en
the $q^2$-dependent width has the structure of the \pipi\ decay rate. 
\\
On the other hand, the $q^2$ dependence for the couplings $g^\jp_{\eta\gamma}$ and $g^\jp_{\epg}$ can be inferred by the QCD structure of the \jp\ radiative decay. Indeed, since the main contribution to this decay is due to the two-gluon intermediate states,  whose Feynman diagram is shown in Fig.~\ref{fig:jpeta}, the coupling should scale as $(\alpha_s(k^2)/k^2)^2$, i.e., as the product of two gluon propagators, being $k$ the gluon four-momentum.
\\
As a consequence, following the same procedure of Eq.~\eqref{eq:pipi-q2}, the couplings $g^\jp_{\eta\gamma}$ and $g^\jp_{\epg}$ have to be substituted as 
\be
g_{h\gamma}^{J/\psi}\hh\to\hh 
g_{h\gamma}^{J/\psi}\, \left(\frac{M_{\jp}^2}{q^2}\right)^2 
\,,\hh\hh h=\eta,\eta'\,.
\label{eq:hg-q2}
\en
\begin{figure}[h!]
\begin{center}
\includegraphics[height=25 mm]{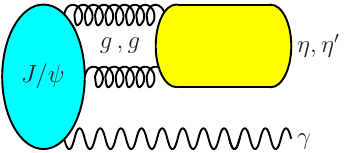}	
\caption{\label{fig:jpeta}Feynman diagram of the radiative decay $\jp\to\eta^{(\prime)}\gamma$. The color scheme is the same of Fig.~\ref{fig:ozi2}.}
\end{center}
\end{figure}\\
Using the definitions of Eqs.~\eqref{eq:pipi-q2} and~\eqref{eq:hg-q2}, the $q^2$ dependent imaginary part of $\agg^{\eta+\eta'}$ reads
\be
  \im[\agg^{\eta+\eta'}(q^2)]&=&{\epsilon_3^\jp\!\sqrt{q^2\! -\! 4M_{\pi}^2} \over 96 \pi \,q^2} 
  \,\frac{M_\jp^4}{(q^2)^2}\,\frac{|D_\rho(M_\jp^2)|}{M_\jp^2\!-\!M_{\rho}^2}
  \label{eq:im-agg-q2}\,\,\,\,
 \\
&&\times\frac{\ds\sum_{h=\eta,\eta'}g_{h\gamma}^{\pi \pi} g_{h \gamma}^{J/\psi} \! \left( q^2-M_{h}^2 \right)^{3}}{ \sqrt{\big(q^2 - M_{\rho}^2\big)^{\! 2} + \Gamma_\rho^2 M_\rho^2 \Big( {q^2-4 M_\pi^2 \over M_{\rho}^2-4 M_\pi^2} \Big)^3 }}\,.\,\,\,\,\,
\nen
The real part is obtained using the DR of Eq.~\eqref{eq:DR-re1}.
\\
In order to account for systematic effects related to the form of $\rho^0$ propagator, besides that of Eq.~\eqref{eq:flatte}, also the propagator with constant width is considered. Hence, in terms of the inverse propagator, the two cases are 
\be
&\ds
D^j_\rho(q^2)=q^2-M_\rho^2+iM_\rho\,\left\{\begin{array}{lcl}
\Gamma_\rho && j=0 \\
\Gamma_\rho \left(\frac{q^2-4M_\pi^2}{M_\rho^2-4M_\pi^2}\right)^{3/2}&& j=1\\
\end{array}
\right.\,,&\no\\\
&&\label{eq:2-drho}
\en
with the corresponding real parts
\be
\re[\agg^{j}(M_\jp^2)]&=&
\frac{\epsilon_3^\jp \!M_\jp^4}{ 96 \pi^2} 
  \,\frac{|D^j_\rho(M_\jp^2)|}{M_\jp^2\!-\!M_{\rho}^2}
\!\sum_{h=\eta,\eta'}\!\!g_{h\gamma}^{\pi \pi} g_{h \gamma}^{J/\psi} 
\no\\
&&\times\!\Pr\!\! \int_{4M^2_\pi}^\infty \!\!\!
  \frac{\left( q^2\!-\!M_{h}^2 \right)^{3}\!\!\sqrt{q^2 \!-\! 4M_{\pi}^2} \,dq^2}{ (q^2\!-\!M_\jp^2)(q^2)^3|D^j_\rho(q^2)|}\,.
\nen
The ratios between real and imaginary part in the two cases $j=0,1$ are 
\be
\begin{array}{lcr}
\ds\frac{\re[\agg^{0}(M_\jp^2)]}{\im[\agg(M_\jp^2)]}&=& 1.910\pm 0.076\,,\vspace{3mm}\\
\ds\frac{\re[\agg^{1}(M_\jp^2)]}{\im[\agg(M_\jp^2)]}&=& 2.096\pm 0.087\,.\\
\end{array}
\label{eq:re-01}
\en
The errors have been propagated by means of a Monte Carlo procedure\footnote{%
%
%
The quantity to be computed $V$, in our case $V=\re(\agg)$, depends on a set of $n$ measured parameters $\{p_j\pm\delta p_j\}_{j=1}^n$, i.e., $V=V(p_1,p_2,\ldots,p_n)$. Starting from such a set, $N$ sets, $\{p_j^{(k)}\pm\delta p_j\}_{j=1}^n$, with $k=1,2,\ldots,N$, are generated by Gaussian fluctuations, so that we have the set $\{V_k=V (p_1^{(k)},p_2^{(k)},\ldots,p_n^{(k)})\}_{k=1}^N$ of the corresponding values for $V$. The final result is 
$$V=\overline V\pm\delta V\,,\hh\hh
\overline V=\sum_{k=1}^N \frac{V_k}{N}\,,\hh\hh
(\delta V)^2=\sum_{k=1}^N\frac{\left(\overline V-V_k\right)^2}{N-1}\,.
$$
}.
Combining the results of Eq.~\eqref{eq:re-01} we obtain
\be
\frac{\re[\agg^{\eta+\eta'}(M_\jp^2)]}{\im[\agg^{\eta+\eta'}(M_\jp^2)]} = 2.00 \pm 0.07^{\rm stat} \pm 0.09^{\rm sys} \,.
\label{eq:re/im}
\en
The statistical error results from the propagation of the two errors obtained by means of the Monte Carlo procedure applied on the two cases $j=0$ and $j=1$, while the systematic error is the half difference of the values given in Eq.~\eqref{eq:re-01}.
\\ 
In light of this result and using Eq.~\eqref{eq:b-gg} to sum up the contributions and Eq.~\eqref{eq:bim-gg}, where $\bgg^{\im}$ and, hence, $\bgg^{\re}$ are proportional to the mean squared value of $\im(\agg)$ and $\re(\agg)$, respectively, the total $\bgg(\eta+\eta')$ BR due to the pseudoscalar contributions is
\be
\bgg^{\eta+\eta'}(\pipi) &=&\left(1+\frac{\overline{(\re[\agg^{\eta+\eta'}(M_\jp^2)])^2}}{\overline{(\im[\agg^{\eta+\eta'}(M_\jp^2)])^2}}\right)\bgg^{\im}(\eta+\eta')
\no\\
&=&(5.78 \pm 0.45^{\rm stat} \pm 0.43^{\rm sys}) \times 10^{-5}\,,
\label{eq:bgg-tot}
\en
where we have used the value of Eq.~\eqref{eq:bgg-value} for $\bgg^{\im}(\eta+\eta')$. 
Finally, by considering the electromagnetic contribution of Eq.~\eqref{eq:bg-pipi}, the total BR from Eq.~\eqref{eq:bpi-2amp}, \hl{still considering the only pseudoscalar contributions}, is
\be
\b^{\eta+\eta'}(\pipi)&=&
(11.4 \pm 2.0^{\rm stat} \pm 0.4^{\rm sys}) \times 10^{-5} 
\no\\
&& + \mathcal{I}^{\eta+\eta'}(\pipi) \,.
\nen
The interference term can be written as 
\be
\mathcal{I}^{\eta+\eta'}(\pipi)=2\sqrt{\bg(\pipi)\bgg^{\eta+\eta'}(\pipi)}\cos(\varphi)\,,
\nen
where, as defined in Eq.~\eqref{eq:rel-phase}, $\varphi$ is the relative phase between the amplitudes \ag\ and $\agg^{\eta+\eta'}$. Since the two BRs have similar values, i.e., $|\ag|\simeq|\agg^{\eta+\eta'}|$, the interference can 
play an important role, indeed, having 
\be
2\sqrt{\bg(\pipi)\bgg^{\eta+\eta'}(\pipi)}= (11.4\pm 2.0)\times 10^{-5}\,,\no \\
\label{eq:int-term}
\en
(statistical and systematic errors have been summed in quadrature) in case of constructive interference, $\varphi=0$, it can even double the effect due to the sum of the single contributions, $\bg+\bgg^{\eta+\eta'}$, on the other hand, it can also cancel out such contributions, in case of destructive interference, $\varphi=\pi$.
\\
From the knowledge of the real and imaginary parts of $\agg^{\eta+\eta'}$, their ratio is given in Eq.~\eqref{eq:re/im}, we may compute the absolute phase of $\agg^{\eta+\eta'}$ as
\be
\phi_{gg\gamma}=\arctan\left(\frac{\im(\agg^{\eta+\eta'})}{\re(\agg^{\eta+\eta'})}\right)
=0.46\pm 0.03 = 13^\circ \pm 1^\circ\,.
\nen
%
%
%
%


\begin{thebibliography}{}
%
%
%
\bibitem{Kopke:1988cs}
\hl{L.~Kopke and N.~Wermes,
  Phys.\ Rept.\  {\bf 174} (1989) 67.}
%
  \bibitem{Milana:1993wk}
  J.~Milana, S.~Nussinov and M.~G.~Olsson,
  Phys.\ Rev.\ Lett.\  {\bf 71} (1993) 2533
  doi:10.1103/PhysRevLett.71.2533
  [hep-ph/9307233].
%
%
\bibitem{Ferroli:2016zqf}
  R.~B.~Ferroli {\it et al.},
  Phys.\ Rev.\ D {\bf 95} (2017) no.3,  034038
  [arXiv:1608.07191 [hep-ph]].
%

%
\bibitem{babar-4pi}
B. Aubert {\it et al.} [\textsc{BaBar} Collaboration], Phys. Rev. D {\bf 71} (2005) 052001 [hep-ex/0502025];
J. P. Lees {\it et al.} [\textsc{BaBar} Collaboration], Phys. Rev. D {\bf 85} (2012) 112009 [arXiv:1201.5677 [hep-ex]].
%
\bibitem{babar-6pi}
B. Aubert {\it et al.} [\textsc{BaBar} Collaboration], Phys. Rev. D {\bf 73} (2006) 052003 [hep-ex/0602006].
%
\bibitem{pdg}
  K.~A.~Olive {\it et al.} [Particle Data Group Collaboration],
  Chin.\ Phys.\ C {\bf 38} (2014) 090001.
%
%
\bibitem{Lees:2012cj}
  J.~P.~Lees {\it et al.} [\textsc{BaBar} Collaboration],
  Phys.\ Rev.\ D {\bf 86} (2012) 032013
  [arXiv:1205.2228 [hep-ex]].
%
%
\bibitem{Gounaris:1968mw}
  G.~J.~Gounaris and J.~J.~Sakurai,
  Phys.\ Rev.\ Lett.\  {\bf 21} (1968) 244.
%
%
\bibitem{Czyz:2009vj}
\hl{  H.~Czyz and J.~H.~Kuhn,
  Phys.\ Rev.\ D {\bf 80} (2009) 034035
  [arXiv:0904.0515 [hep-ph]].}
%
\bibitem{Seth:2012nn}
\hl{  K.~K.~Seth, S.~Dobbs, Z.~Metreveli, A.~Tomaradze, T.~Xiao and G.~Bonvicini,
  Phys.\ Rev.\ Lett.\  {\bf 110} (2013) no.2,  022002
  [arXiv:1210.1596 [hep-ex]].}
%
\bibitem{cut}
  R.~E.~Cutkosky,
  J.\ Math.\ Phys.\  {\bf 1} (1960) 429.
%
\bibitem{Field:2001iu}
 \hl{ J.~H.~Field,
  Phys.\ Rev.\ D {\bf 66} (2002) 013013
  [hep-ph/0101158].}
%
\bibitem{Pacetti:2009pg}
  S.~Pacetti,
  Nucl.\ Phys.\ A {\bf 919} (2013) 15
  [arXiv:0904.3684 [hep-ph]].
%
\hl{\bibitem{Rudenko:2017bel}
  A.~S.~Rudenko,
  Phys.\ Rev.\ D {\bf 96} (2017) no.7,  076004
  [arXiv:1707.00545 [hep-ph]];
  J.~H.~Kuhn, J.~Kaplan and E.~G.~O.~Safiani,
  Nucl.\ Phys.\ B {\bf 157} (1979) 125;
  G.~Kopp, T.~F.~Walsh and P.~M.~Zerwas,
  Nucl.\ Phys.\ B {\bf 70} (1974) 461;
  F.~M.~Renard,
  Nuovo Cim.\ A {\bf 80} (1984) 1.}
%
\bibitem{flatte}
  S.~M.~Flatt\'e,
  Phys.\ Lett.\ B {\bf 63} (1976) 224.

%
%
\end{thebibliography}
\end{document}